\documentclass[aps,prl,reprint,graphicx,footinbib,floatfix,superscriptaddress]{revtex4-1}

\usepackage{amsmath}
\usepackage[pdftex]{graphicx}
\usepackage{hyperref}
\usepackage[english]{babel}
\usepackage{units}
\usepackage{braket}
\usepackage{tabularx}
\usepackage{dcolumn}
\usepackage{units}
\usepackage{multirow}
\newcolumntype{d}[1]{D{.}{.}{#1}}

\hypersetup{
	colorlinks=true,
	allcolors=black,
	breaklinks=true}

\newcommand{\affA}{Physikalisch-Technische Bundesanstalt, Bundesallee 100, 38116
	Braunschweig, Germany} 
\newcommand{\affB}{Institut f\"ur Quantenoptik, Leibniz
	Universit\"at Hannover, Welfengarten 1, 30167 Hannover, Germany}
\newcommand{\affC}{Laboratory for Nano and Quantum Engineering, Leibniz Universit{\"a}t Hannover, Schneiderberg 39, 30167 Hannover, Germany}

\newcommand{\thetitle}{$^{115}$In$^+$-$^{172}$Yb$^+$ Coulomb crystal clock with $2.5\times10^{-18}$ systematic uncertainty}
\newcommand{\thedate}{October 25, 2024}

\begin{document}	
	\title{\thetitle}
	
	\author{H.~N.~Hausser} 
	\thanks{These authors contributed equally to this work.}
	\affiliation{\affA}
	
	\author{J.~Keller} 
	\thanks{These authors contributed equally to this work.}
	\affiliation{\affA}	
	
	\author{T.~Nordmann} 
	\affiliation{\affA}
	\author{N.~M. Bhatt} 
	\affiliation{\affA}
	\author{J.~Kiethe} 
	\affiliation{\affA}
	\author{H.~Liu} 
	\affiliation{\affA}
	\author{I.~M.~Richter} 
	\affiliation{\affA}
	\author{M.~von~Boehn} 
	\affiliation{\affA}
	\author{J.~Rahm} 
	\affiliation{\affA}
	\author{S.~Weyers} 
	\affiliation{\affA}
	\author{E.~Benkler} 
	\affiliation{\affA}
	\author{B.~Lipphardt} 
	\affiliation{\affA}
	\author{S.~D{\"o}rscher} 
	\affiliation{\affA}
	\author{K.~Stahl} 
	\affiliation{\affA}
	\author{J.~Klose} 
	\affiliation{\affA}
	\author{C.~Lisdat} 
	\affiliation{\affA}
	\author{M.~Filzinger} 
	\affiliation{\affA}
	\author{N.~Huntemann} 
	\affiliation{\affA}
	\author{E.~Peik} 
	\affiliation{\affA}
	\author{T.~E.~Mehlst{\"a}ubler} 
	\email[]{tanja.mehlstaeubler@ptb.de}
	\affiliation{\affA}
	\affiliation{\affB}
	\affiliation{\affC}

	\date{\thedate}

	\begin{abstract}
		We present a scalable mixed-species Coulomb crystal clock based on the $^1S_0$ $\leftrightarrow$ $^3P_0$ transition in $^{115}$In$^+$. $^{172}$Yb$^+$ ions are co-trapped and used for sympathetic cooling. Reproducible interrogation conditions for mixed-species Coulomb crystals are ensured by a conditional preparation sequence with permutation control. We demonstrate clock operation with a 1In$^+$--3Yb$^+$ crystal, achieving a relative systematic uncertainty of $2.5\times10^{-18}$ and a relative frequency instability of $1.6\times10^{-15}/\sqrt{\tau/\unit[1]{s}}$. We report on absolute frequency measurements with an uncertainty of $1.3\times10^{-16}$ and optical frequency comparisons with clocks based on $^{171}$Yb$^+$ (E3) and $^{87}$Sr. With a fractional uncertainty of $4.4\times10^{-18}$, the former is -- to our knowledge -- the most accurate frequency ratio value reported to date. For the $^{115}$In$^+$/$^{87}$Sr ratio, we improve upon the best previous measurement by more than an order of magnitude. We also demonstrate operation with four $^{115}$In$^+$ clock ions, which reduces the instability to $9.2\times10^{-16}/\sqrt{\tau/\unit[1]{s}}$.
	\end{abstract}
	
	\maketitle

    Optical clocks with $10^{-18}$ level fractional frequency uncertainties \cite{Ludlow2015} enable precise tests of fundamental physics \cite{Safronova2018}, new applications such as chronometric leveling \cite{Lion2017, Mehlstaeubler2018, Grotti2018} and are the prerequisite for a future redefinition of the SI unit of time \cite{Dimarcq2024}. In order to validate their performance, repeated comparisons between independent systems, operated by separate laboratories and employing different species, are necessary at this level \cite{Doerscher2021, BACONC2021, Kim2022}. To this day, frequency ratios of different optical transitions have been measured with uncertainties as low as $5.9\times10^{-18}$ \cite{BACONC2021}. A very promising species for low systematic uncertainties is $^{115}$In$^+$ \cite{Dehmelt1982,Becker2001, Wang2007, Ohtsubo2017, Ohtsubo2020}. In 2012, In$^+$ was proposed as a favorable candidate for a multi-ion clock \cite{Herschbach2012}, addressing a fundamental problem in trapped-ion optical clocks: As their systematic uncertainties are reduced, measurements are increasingly limited by the statistical uncertainty due to quantum projection noise (QPN) of a single particle \cite{Itano1993, Peik2005}. Multi-ion clocks will open up the path for ion clock measurements with $10^{-19}$ level overall uncertainties \cite{Keller2019} or can relax local oscillator stability requirements \cite{Keller2014}.
	
    In this Letter, we demonstrate the operation of an $^{115}$In$^+$ clock which is based on linear Coulomb crystals (CC) and can be operated with a variable number of clock ions. We first evaluate our new clock setup with a CC containing a single In$^+$ ion, obtaining a fractional systematic uncertainty of $2.5\times10^{-18}$. In comparisons with two other optical clocks, we observe averaging behavior compatible with white frequency noise down to this level. We report measurements of the frequency ratios with respect to $^{87}$Sr and the electric octupole (E3) transition in $^{171}$Yb$^+$, with respective relative uncertainties of $4.2\times10^{-17}$ -- more than one order of magnitude lower than previously reported \cite{Ohtsubo2020} -- and $4.4\times10^{-18}$. In addition, we determine the absolute frequency of the $^1S_0$ $\leftrightarrow$ $^3P_0$ transition in $^{115}$In$^+$ with an uncertainty limited by the realization of the SI second, which constitutes a further requirement for the optical redefinition of the SI second \cite{Dimarcq2024}. Finally, we demonstrate scalability by operation with up to four clock ions.
	
	In our setup, $^{115}$In$^+$ ions are co-trapped with $^{172}$Yb$^+$ ions, which provide sympathetic cooling on a strong, dipole-allowed transition ($\Gamma\approx2\pi\times\unit[20]{MHz}$). Their positions within the crystal determine the cooling rates of its motional degrees of freedom \cite{Kielpinski2000, Home2013}. Since background gas collisions can enable swapping of ion positions, we have developed a mechanism to reproducibly restore crystals to a target permutation, which we identify by a binary string in this work (0$\widehat{=}$In$^+$, 1$\widehat{=}$Yb$^+$). For example, with the 1In$^+$--3Yb$^+$ composition used below, we operate the clock with the permutation 1011 (or its mirror image), for which the motional modes with predominant In$^+$ ion motion have four times higher sympathetic cooling rates than those of the permutation 0111. This ensures reproducible and low kinetic energies with a fixed duration cooling pulse.
 
    The CC is trapped in a scalable 3D chip ion trap \cite{Keller2019}. The secular frequencies for $^{172}$Yb$^+$ center-of-mass (COM) motion are ($\omega_{\text{rad1}}, \omega_{\text{rad2}}, \omega_{\text{ax}})/2\pi\approx(\unit[822]{kHz}, \unit[794]{kHz},\unit[275]{kHz})$. A bichromatic imaging system allows site-resolved fluorescence detection of both species on an EMCCD camera via the $^2S_{1/2}$ $\leftrightarrow$ $^2P_{1/2}$ transition at \unit[369.5]{nm} for the $^{172}$Yb$^+$ ions and the $\ket{^1S_0, F=9/2, m_F=\pm9/2} \leftrightarrow \ket{^3P_1, F=11/2, m_F=\pm11/2}$ transitions at \unit[230.6]{nm} for the $^{115}$In$^+$ ions \cite{Nordmann2023}. The light used to interrogate the clock transition is derived from a high-finesse-cavity-stabilized \unit[946]{nm} Nd:YAG laser \cite{Didier2019} which is transfer-locked to an ultrastable cryogenic silicon resonator \cite{Matei2017}. Further details of the experimental setup can be found in Ref.~\cite{Nordmann2023a}.
		
	\begin{figure}
		\includegraphics[width=\columnwidth]{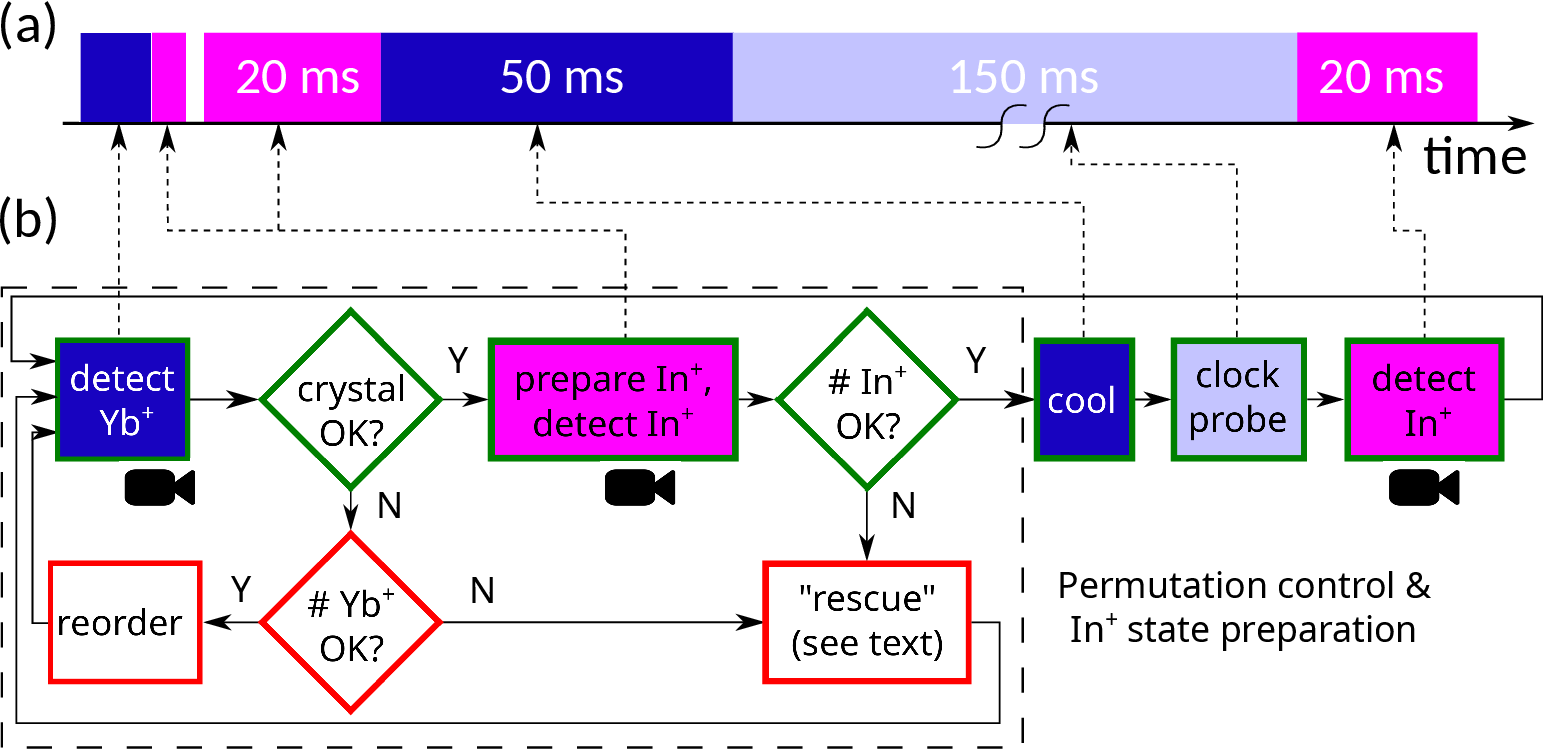}
		\caption{Coulomb crystal clock cycle sequence. a: time sequence, b: decision chart. The crystal is initialized in the target permutation with conditional reordering, cooling and molecule dissociation steps. Rabi interrogation is carried out after sympathetic cooling on the Yb$^+$ ions and followed by state detection via electron shelving on the In$^+$ detection transition.}
		\label{fig:clock_cycle}
	\end{figure}

	The clock interrogation loop used to stabilize the laser frequency to the atomic reference line is shown in Fig.~\ref{fig:clock_cycle}. Each iteration consists of four stages: preparation, cooling, interrogation, and detection. In the preparation stage, the system determines the position of the ytterbium ions in the crystal via \unit[369.5]{nm} fluorescence, reorders the ions if necessary and optically pumps the indium ions into either of the $\ket{^1S_0, m_\text{F}=\pm9/2}$ states. A \unit[230.6]{nm} fluorescence measurement verifies the initialization of all clock ions. If too few ions of either species are detected, additional cooling laser frequency sweeps to recrystallize and illumination with about $\unit[10]{\mu W}$ at $\unit[230.6]{nm}$ focused to $\unit[90]{\mu m}$ to dissociate molecular ions -- likely YbOH$^+$ formed in background gas collisions -- are applied (``rescue'' branch in Fig.~\ref{fig:clock_cycle}b). After successful preparation, the crystal is Doppler cooled via the Yb$^+$ ions. A rectangular pulse of \unit[150]{ms} duration probes the clock transition. Finally, a fluorescence measurement determines the indium ions' states. A measurement is discarded if the crystal is found in a different permutation in the next iteration. The entire clock sequence is thus ready for operation with a linear CC of multiple clock and cooling ions. We demonstrate this concept by using a 1In$^+$--3Yb$^+$ CC for the measurements presented in the following sections.
	
	We apply a bias magnetic field of about \unit[106]{$\mu$T}, which is more than an order of magnitude higher than typical magnetic field fluctuations in our setup, along the axial trap direction and interrogate the two Zeeman transitions $\ket{^1S_0, m_F=\pm9/2} \leftrightarrow \ket{^3P_0, m_F=\pm9/2}$ (split by approximately \unit[4280]{Hz}) using $\pi$-polarized light at \unit[236.5]{nm}. Averaging the two components provides a first-order Zeeman shift insensitive frequency measurement and in-situ magnetic field determination. We observe a Fourier limited linewidth and a contrast of about \unit[60]{\%}, which is consistent with the excited state lifetime of \unit[195(8)]{ms} \cite{Becker2001} and residual thermal motion after sympathetic cooling.
		
	\begin{table}
		\caption{Fractional frequency shifts and associated uncertainties for the $^{115}$In$^+$ Coulomb crystal clock operated in 1In$^+$--3Yb$^+$ composition. Further contributions with uncertainties below $2\times10^{-19}$ are omitted \cite{Supplemental}.}
		\begin{tabularx}{\columnwidth}{l d{3.2} d{3.2}}
			\hline
			
		\multicolumn{1}{l}{Effect} \rule{0pt}{3ex} & \multicolumn{1}{c}{Shift (10$^{-18}$)} & \multicolumn{1}{c}{Uncertainty (10$^{-18}$)}  \\
			\hline
			\rule{0pt}{3ex}Thermal time dilation & -2.7 & 1.6 \\ 
			Black-body radiation & -13.4 & 1.4 \\ 
			Quadratic Zeeman & -35.9 & 1.1 \\ 
			Servo Error & -2.6 & 0.5 \\ 
            AOM chirp & -0.5 & 0.5 \\
			Background gas collisions & 0 & 0.4 \\ 
			Time dilation (EMM) & -0.8 & 0.1 \\ 
			Electric quadrupole & -0.14 & 0.03 \\
			\rule{0pt}{3ex}Total & -56.0 & 2.5 \\ 
			\hline
		\end{tabularx}
		\label{tab:uncertainty_budget}
	\end{table}	
	
   	The systematic shifts of the clock operating with a 1In$^+$--3Yb$^+$ CC in permutation 1011 and their uncertainties are summarized in Tab.~\ref{tab:uncertainty_budget}. The overall fractional systematic uncertainty is evaluated to be $u_\mathrm{B}^{\mathrm{In^+}}=2.5\times10^{-18}$. In the following, the individual contributions are discussed in order of relevance. Additional details for all contributions can be found in the Supplemental Material \cite{Supplemental}.
	
	Thermal time dilation (TD) results from the residual motion of the clock ion after sympathetic cooling and the corresponding intrinsic micromotion \cite{Keller2019}. Due to the low heating rates in our trap ($<\unit[1]{s^{-1}}$ for the radial Yb$^+$ COM modes) \cite{Nordmann2023a}, we assume constant crystal temperatures throughout the clock interrogation. The radial temperatures are determined from thermal dephasing of Rabi oscillations and used to calibrate our sympathetic cooling model, from which we derive the axial temperatures. This yields kinetic energy bounds of $k_B\times\left( T_{\text{rad1}} + T_{\text{rad2}}\right)= k_B\times\unit[2.3(14)]{mK}$ for the radial modes with the lowest cooling rates and $k_B\times\left( T_{\text{rad1}} + T_{\text{rad2}}\right)= k_B \times\unit[0.9(1)]{mK}$ each for all other radial modes. Together with the axial temperatures, which are estimated as $T_{\text{ax}}=\unit[1.7(10)]{mK}$ for all modes, we obtain a total shift of $\Delta\nu_\mathrm{TD}/\nu_0=-2.7(16)\times10^{-18}$. The temperature uncertainty is currently limited by the uncertainty of the angle between clock laser k-vector and radial mode principal axes $\theta_1 = \unit[40(13)]{^\circ} = \unit[90]{^\circ}-\theta_2$. This angle will be determined more accurately in future sideband spectroscopy measurements. Directly cooling the clock ions on the $^1$S$_0$$\leftrightarrow$$^3$P$_1$ intercombination line ($\Gamma=2\pi\times\unit[360]{kHz}$) will yield overall thermal TD shifts $<2\times10^{-19}$ \cite{Peik1999, Kulosa2023}. 
	
	The black-body radiation (BBR) shift is calculated based on the static differential polarizability $\Delta\alpha_\mathrm{stat}=\unit[3.3(3)\times10^{-41}]{J/(V/m)^2}$ \cite{Safronova2011} of the clock transition. At room temperature, the shift uncertainty is dominated by the uncertainty of this value. We therefore assume the temperature uncertainty of the ion's environment to be half of the entire span of temperatures observed during the clock uptime from sensors on the trap \cite{Nordmann2020} and chamber, with $T=\unit[299(1)]{K}$, as this translates to a negligible $2\times10^{-19}$ uncertainty contribution to the total shift of $\Delta\nu_\mathrm{BBR}/\nu_0=-13.4(14)\times10^{-18}$. The temperature contribution could be further reduced with a time-resolved instead of constant correction.

	The quadratic Zeeman shift is calculated as $\Delta\nu=\beta\langle B^2 \rangle $ with the coefficient $\beta=\unit[-4.05]{Hz/mT^2}$~\cite{Herschbach2012}\footnote{Note that the value is stated in \cite{Herschbach2012} without sign.}. The time-resolved magnetic field data, determined from the observed Zeeman splitting, are used to post-correct the measured frequencies. The uncertainty in this magnetic field data is dominated by the knowledge of the excited state g-factor $g(^3P_0) = -9.87(5)\times10^{-4}$ \cite{Becker2001}. The average shift is determined to be $\Delta\nu_\mathrm{Z2}/\nu_0=-35.9(11)\times10^{-18}$. The uncertainty can be reduced with a more precise measurement of $g(^3P_0)$. For the AC contribution due to trap rf currents, simulations indicate $B^2_{\text{rms}}=\unit[1.3\times10^{-12}]{T^2}$ \cite{Dolezal}, which is comparable to experimental observations in other ion traps \cite{Gan2018, Brewer2019}. Even the highest reported value to date of $B^2_{\text{rms}}=\unit[2.17\times10^{-11}]{T^2}$ \cite{Chou2010} amounts only to a fractional frequency shift of $-7\times10^{-20}$ in $^{115}$In$^+$.

	The transfer lock reduces the linear drift rate of the clock laser to that of the Si cavity of ca.  $\unit[-120]{\mu Hz\;s^{-1}}$. Due to an imperfect compensation algorithm, this residual drift resulted in a servo offset of $\Delta\nu_\mathrm{servo}/\nu_0 = -2.6(5)\times10^{-18}$, as determined in post-processing using the logged error signal \cite{Supplemental}.

    Thermal effects in the clock laser AOM can cause cycle-synchronous phase chirps which result in a systematic frequency offset. From interferometric phase measurements \cite{Kazda2016} at different drive powers, we infer a fractional shift of $-5(5)\times10^{-19}$.
 
	Site-resolved detection in mixed-species operation allows us to detect all background gas collisions which result in CC permutation changes. In the 1In$^+$--3Yb$^+$ composition, this corresponds to \unit[75]{\%} of all events. Collisions below the ion-swapping energy barrier constitute a negligible fraction of events \cite{Supplemental}. Thus, from a measured collision rate per ion of $\Gamma_{\text{meas}}^{\text{ion}}=\unit[0.0029(3)]{s^{-1}}$ \cite{Nordmann2023a}, we can estimate the rate of undetected collisions for the four-ion crystal as $\Gamma_{\text{undet}}^{\text{CC}}=4\times0.25\times\Gamma_{\text{meas}}^{\text{ion}}=\unit[0.0029(3)]{s^{-1}}$. The treatment of Ref.~\cite{Rosenband2008} for the collision-induced phase shifts yields a fractional frequency shift of $\Delta\nu_\mathrm{coll}/\nu_0 = 0(4)\times10^{-19}$. TD shifts due to transferred kinetic energy which would be relevant at this level are suppressed by their detrimental effect on state detection \cite{Hankin2019, Supplemental}.
	
	Excess micromotion (EMM) leads to additional TD and Stark shifts \cite{Berkeland1998}, the inhomogeneity of which is below $1\times10^{-19}$ across the crystal \cite{Keller2019a, Nordmann2023a}. We compensate radial stray fields via photon-correlation measurements using the Yb$^+$ ions \cite{Keller2015} at least once every eight hours. The time-averaged residual shift is calculated in post-processing by linearly interpolating electric stray fields between \unit[0]{V/m} and the value observed in the subsequent compensation measurement. Axial EMM contributes $\Delta\nu_\mathrm{EMM}^\mathrm{(ax)}/\nu_0=-8(1)\times10^{-19}$ and determines the overall shift. 
 
    The electric quadrupole shift is calculated via the Hamiltonian $H_\mathrm{E2} = \nabla E^\text{(2)}\Theta^\text{(2)}$ \cite{Itano2000} with the E-field gradient $\nabla E^\text{(2)}$ accounting for the trap electric field and Coulomb interaction within the ion crystal \cite{Keller2019} and the electric quadrupole moment $\Theta = -1.6(3)\times10^{-5}ea_B^2$ \cite{Beloy2017}, yielding $\Delta\nu_\mathrm{E2}/\nu_0=-1.4(3)\times10^{-19}$.
	
	Further shifts, such as the clock laser AC Stark and first-order Doppler shifts have been estimated to contribute less than $2\times10^{-19}$ to the overall uncertainty \cite{Supplemental}.

	\begin{figure}
		\includegraphics[width=\columnwidth]{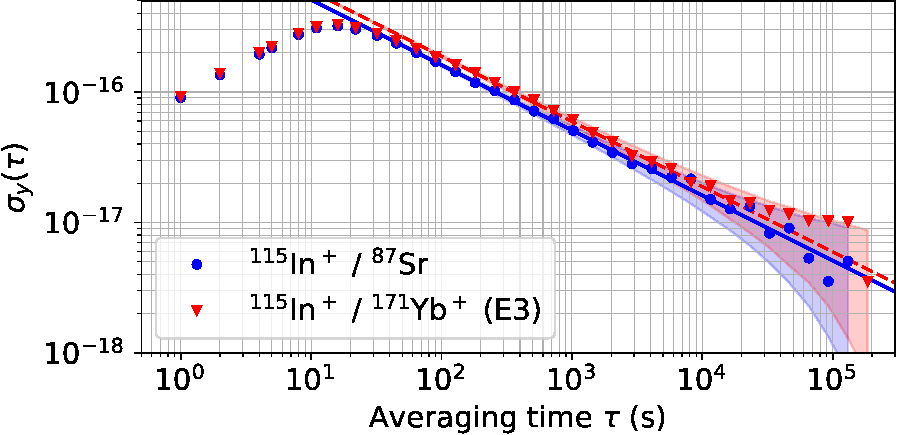}
		\caption{Instabilities (ADEV) of the $^{115}$In$^+$/$^{87}$Sr (blue circles) and $^{115}$In$^+$/$^{171}$Yb$^+$ (E3) (red triangles) frequency ratio measurements. The fitted instabilities are $\sigma_{\text{y}}^{\mathrm{fit, In^+/Sr}}=1.6\times10^{-15}/\sqrt{\tau/\unit[1]{s}}$ (solid blue line) and $\sigma_{\text{y}}^{\mathrm{fit, In^+/Yb^+}}=1.9\times10^{-15}/\sqrt{\tau/\unit[1]{s}}$ (dashed red line). The shaded areas represent 95\% confidence intervals for white frequency noise datasets of the respective durations, as determined in Monte Carlo simulations.}
		\label{fig:In_vs_Sr_Yb_adev}
	\end{figure}
	
	We have compared the $^{115}$In$^+$ clock to an $^{171}$Yb$^+$ (E3) single-ion clock \footnote{``clock 1'' in Ref.~\cite{Sanner2019}} and $^{87}$Sr lattice clock \cite{Schwarz2022} via an optical frequency comb. In both cases, the measurement instability reaches a few parts in $10^{18}$ for total measurement times of about one week (see Fig.~\ref{fig:In_vs_Sr_Yb_adev}). The instability of $1.6\times10^{-15}\sqrt{\tau/\unit[1]{s}}$ of the In$^+$/Sr comparison reflects that of the In$^+$ CC clock, since the contribution from the Sr clock is only about $2\times10^{-16}/\sqrt{\tau/\unit[1]{s}}$ \cite{Schwarz2020}. This is consistent with the instability of the In$^+$/Yb$^+$ comparison, given the $^{171}$Yb$^+$ (E3) clock instability of $1.0\times10^{-15}/\sqrt{\tau/\unit[1]{s}}$ \cite{Sanner2019}.
        
	\begin{table*}[ht]
	  \caption{Frequency ratios with combined statistical ($u_\mathrm{A}$), systematic ($u_\mathrm{B1,2}$, where $1$ refers to the In$^+$ clock) and total ($u_\mathrm{C}$) uncertainties (in units of $10^{-18}$), and inferred absolute frequencies for the $^1S_0$ $\leftrightarrow$ $^3P_0$ transition in $^{115}$In$^+$. The absolute frequencies inferred from the optical ratios use results reported in Ref.~\cite{Schwarz2020} (Sr) and Ref.~\cite{Lange2021} (Yb$^+$). All values are corrected for the relativistic redshifts (RRS) due to clock height differences \cite{Supplemental}.}
	  \label{tab:meas_overview}
          \centering
	  \begin{tabularx}{2\columnwidth}{p{0.35\columnwidth}p{0.1\columnwidth}p{0.1\columnwidth}p{0.13\columnwidth}p{0.12\columnwidth}p{0.1\columnwidth}p{0.49\columnwidth}p{0.55\columnwidth}}
	 		\hline	 		
	 	        \multicolumn{1}{l}{Clocks} \rule{0pt}{2ex} & \multicolumn{1}{l}{$u_\mathrm{A}$} & \multicolumn{1}{l}{$u_\mathrm{B1}$} & \multicolumn{1}{l}{$u_\mathrm{B2}$} & \multicolumn{1}{l}{$u_\mathrm{RRS}$} & \multicolumn{1}{l}{$u_\mathrm{C}$} & \multicolumn{1}{l}{Ratio} & \multicolumn{1}{l}{Absolute frequency (Hz)} \\
	 		\hline\rule{0pt}{3ex}
   	 		$^{115}$In$^+$/$^{171}$Yb$^+$ (E3) & 2.4 & 2.5 & 2.7 & 0.5 & 4.4 & 1.973\,773\,591\,557\,215\,789(9) & 1\,267\,402\,452\,901\,038.87(16) \\
            \rule{0pt}{3ex}$^{115}$In$^+$/$^{87}$Sr & 2.4 & 2.5 & 42 & 0.5 & 42 & 2.952\,748\,749\,874\,860\,78(13) & 1\,267\,402\,452\,901\,038.96(21) \rule{0pt}{3ex}\\
	 		$^{115}$In$^+$/$^{133}$Cs & 170$^\ast$ & 15$^{\ast\ast}$ & 170 & 0.7 & 240 & & 1\,267\,402\,452\,901\,039.05(30) \rule{0pt}{3ex}\\
	 		\hline
	 		\rule{0pt}{3ex}CCTF2021 \cite{Margolis2024} & & & & & 4300 & & 1\,267\,402\,452\,901\,041.3(54) \\
	 		\hline 
                        \multicolumn{8}{l}{$^\ast$The combined averaging time of \unit[$6.2\times10^5$]{s} is extended to \unit[$1.0\times10^6$]{s} by using a hydrogen maser as a flywheel for gaps} \\ 
                        \multicolumn{8}{l}{in the combined In$^+$/Cs uptime (cf.~\cite{Schwarz2020, Lange2021}). $u_\mathrm{A}$ contains an extrapolation uncertainty contribution of $u_\mathrm{ext}=9\times10^{-17}$.}\\
                        \multicolumn{8}{l}{$^{\ast\ast}$Additional intervals with $u_\mathrm{B, In^+}$ up to $1.5\times10^{-17}$ due to increased EMM are included in the In$^+$/Cs comparison.}\\
	  \end{tabularx}
	\end{table*}

	 For both datasets, we observe $\chi^2_\mathrm{red}\approx1.3$ \cite{Supplemental} and averaging behaviors compatible with white frequency noise, which correspond to a statistical uncertainty of $u_A=2.4\times10^{-18}$ (in both cases) at the respective dataset duration of $\unit[4.6\times10^5]{s}$ (In$^+$/Sr) and $\unit[6.0\times10^5]{s}$ (In$^+$/Yb$^+$). After correcting for systematic frequency shifts and relativistic shifts from the height difference between the clocks \cite{Supplemental}, we find an optical frequency ratio of $\nu^{\text{In}^+}_0/\nu^{\text{Yb}^+}_0=1.973\,773\,591\,557\,215\,789(9)$.

We find a frequency ratio $\nu^{\text{In}^+}_0/\nu^{\text{Sr}}_{0}=2.952\,748\,749\,874\,860\,78(13)$, where we include a provisional frequency correction estimation which accounts for long-term variations of the Sr clock frequency and correspondingly increases the uncertainty. These variations became apparent in a repeated comparison of the three clocks in 2024, which shows deviations in $\nu^{\text{In}^+}_0/\nu^{\text{Sr}}_0$ and $\nu^{\text{Yb}^+}_0/\nu^{\text{Sr}}_0$ at the mid-$10^{-17}$ level and reproduces the above value for $\nu^{\text{In}^+}_0/\nu^{\text{Yb}^+}_0$within a combined relative uncertainty of $9.7\times10^{-18}$. Details on the repeated measurement and frequency correction are given in the Supplemental Material \cite{Supplemental}. We include the value $\nu^{\text{In}^+}_0/\nu^{\text{Sr}}_{0}$ for comparison with the previously most accurate measurement of the In$^+$ frequency via this ratio \cite{Ohtsubo2020}, with which we find agreement within $1.1\sigma$.
Table \ref{tab:meas_overview} summarizes the measured frequency ratios together with the individual uncertainty contributions.

	 \begin{figure}
	   \includegraphics[width=\columnwidth]{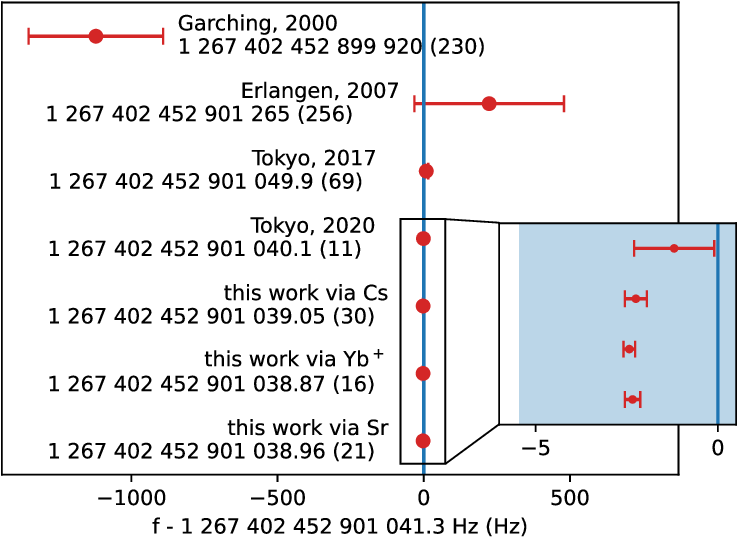}
	   \caption{Measurement history for the unperturbed frequency of the $^1S_0$ $\leftrightarrow$ $^3P_0$ transition in $^{115}$In$^+$ \cite{vonZanthier2000, Wang2007, Ohtsubo2017, Ohtsubo2020}. The blue line and shaded area correspond to the CCTF 2021 recommendation of $f_0~=~\unit[1\,267\,402\,452\,901\,041.3(54)]{Hz}$ \cite{Margolis2024}.}
	   \label{fig:in_absfreq_overview}
	\end{figure}
	The absolute frequency of the unperturbed $^1S_0$ $\leftrightarrow$ $^3P_0$ transition is determined by comparison to PTB's primary Cs fountain clock CSF2 \cite{Weyers2018} via a hydrogen maser, analogous to Refs.~\cite{Schwarz2020, Lange2021}. The optical frequency ratio measurements of Tab.~\ref{tab:meas_overview} provide an indirect method to determine the absolute frequency with reduced statistical uncertainty via previous absolute frequency measurements of $^{171}$Yb$^+$ (E3) \cite{Lange2021} and $^{87}$Sr \cite{Schwarz2020}. The former yields a frequency uncertainty of $u_\mathrm{C}^\mathrm{In^+/Cs}=1.3\times10^{-16}$, limited by the determination of the Yb$^+$ absolute frequency (see Tab.~\ref{tab:meas_overview}). However, it is important to note that these results should not be considered as independent (see \cite{Supplemental} for information regarding correlations). The inferred absolute frequencies are shown in Fig.~\ref{fig:in_absfreq_overview}. They agree well with the recommended frequency value \cite{Margolis2024} as well as previously reported values \cite{Wang2007, Ohtsubo2017, Ohtsubo2020}.
	
	\begin{figure}
		\includegraphics[width=\columnwidth]{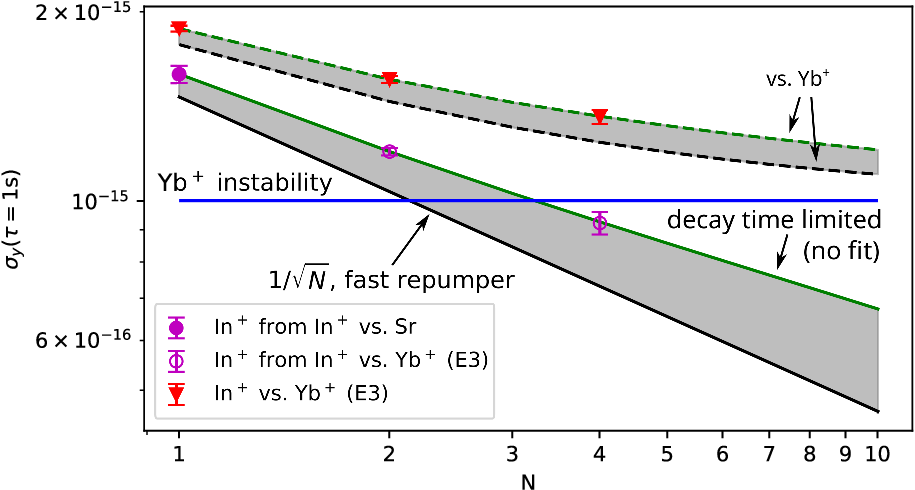}
		\caption{Multi-ion clock frequency instabilities at \unit[1]{s} with up to 4 In$^+$ clock ions. Red triangles depict the combined instability of the In$^+$ / Yb$^+$ ratio. Pink circles show the instability of the In$^+$ clock as extracted from the ratios In$^+$ / Sr (filled) and In$^+$ / Yb$^+$ (open). The solid green line shows the expected QPN-limited scaling with clock ion number, taking into account the additional dead time for state preparation. The addition of a repumper laser for depletion of the $^3P_0$ level would lead to instabilities indicated by the solid black line.}
		\label{fig:multi-ion_instab}
	\end{figure}
	
	The CC clock is implemented with the intent of scaling up the number of clock ions. We demonstrate this capability in instability measurements with up to four clock ions against the $^{171}$Yb$^+$ (E3) clock, as shown in Fig.~\ref{fig:multi-ion_instab}. The chosen compositions, 2In$^+$--4Yb$^+$ and 4In$^+$--8Yb$^+$, are operated in the respective permutations 101011 and 110101010111. We observe reduced instabilities with increasing clock ion number down to $9.2(4)\times10^{-16}$ at \unit[1]{s} for four clock ions. However, state preparation currently relies on the spontaneous decay of the $^3P_0$ level with a lifetime of \unit[195(8)]{ms} \cite{Becker2001} after successful excitation attempts. The associated dead time increases with ion number and reduces the instability advantage of the increased signal. This will be mitigated in future by a quench laser at \unit[481.6]{nm} which returns $^3P_0$ population to the ground state via $^1P_1$.
	
	In summary, we have demonstrated a Coulomb crystal clock based on $^{115}$In$^+$-$^{172}$Yb$^+$ ion chains with a systematic uncertainty of $2.5\times10^{-18}$ and an instability of $1.6\times10^{-15}/\sqrt{\tau/\unit[1]{s}}$ for operation with a 1In$^+$--3Yb$^+$ CC. We have measured the absolute frequency of the $^1S_0$ $\leftrightarrow$ $^3P_0$ transition in $^{115}$In$^+$ as well as the optical frequency ratios $^{115}$In$^+$/$^{87}$Sr and $^{115}$In$^+$/$^{171}$Yb$^+$ (E3) with relative uncertainties of $4.2\times10^{-17}$ and $4.4\times10^{-18}$. To our knowledge, the latter is the frequency ratio measurement with the lowest uncertainty reported to date. Furthermore, we have shown a reduction of the statistical uncertainty to $9.2\times10^{-16}/\sqrt{\tau/\unit[1]{s}}$ in multi-clock-ion operation. This constitutes a significant step towards multi-ion clock operation with low $10^{-16}/\sqrt{\tau/\unit[1]{s}}$ instabilities and $10^{-19}$ level inaccuracies \cite{Keller2019} benefiting future fundamental physics tests and relativistic geodesy.
	
	The presented data will provide information for upcoming adjustments of the CCTF recommended frequency value for the $^{115}$In$^+$ clock transition and contribute to ongoing consistency checks of optical frequency measurements towards a redefinition of the SI unit of time.
        
        Since the major contributions to the uncertainty budget are limited by either atomic constants or our limited knowledge of the radial principal axis orientation $\theta_1$, we expect a current reproducibility of the In$^+$ CC clock of about $6\times10^{-19}$.

\begin{acknowledgments}
	We thank Uwe Sterr, Thomas Legero and Jialiang Yu for providing the ultra-stable \unit[1542]{nm} laser, Stepan Ignatovich and Maksim Okhapkin for their work on the clock laser, Heiner Denker and Ludger Timmen for leveling measurements of our clock systems, Fabian Wolf for support with UV fiber production, and Rattakorn Kaewuam, Piyaphat Phoonthong and Michael Kazda for their help in the determination of the AOM phase chirp.
	
	We acknowledge support by the projects 18SIB05 ROCIT and 20FUN01 TSCAC. These projects have received funding from the EMPIR programme co-financed by the Participating States and from the European Union’s Horizon 2020 research and innovation programme. We acknowledge funding by the Deutsche Forschungsgemeinschaft (DFG) under Germany’s Excellence Strategy – EXC-2123 QuantumFrontiers –390837967 (RU B06) and through Grant No. CRC 1227 (DQ-mat, projects B02 and B03). This work has been supported by the Max-Planck-RIKEN-PTB-Center for Time, Constants and Fundamental Symmetries.

\end{acknowledgments}

\bibliography{references}
\clearpage
\pagebreak

\onecolumngrid
\begin{center}
	\textbf{\large Supplemental Material: \thetitle}
\end{center}
\twocolumngrid
\setcounter{equation}{0}
\setcounter{figure}{0}
\setcounter{table}{0}
\setcounter{page}{1}
\renewcommand{\theequation}{S\arabic{equation}}
\renewcommand{\thefigure}{S\arabic{figure}}
\renewcommand{\thetable}{S\Roman{table}}

\section{Systematic uncertainties}
Below, we provide details on all frequency shifts and systematic uncertainty contributions of the $^{115}$In$^+$ clock, as summarized in Tab.~\ref{tab:uncertainty_budget_detailed}.

\begin{table}[h]
	\caption{\label{tab:uncertainty_budget_detailed}Overview of all contributions to the systematic uncertainty and the different contributions to the individual uncertainties.}
	\begin{tabularx}{\columnwidth}{|l|l|X|}
	\hline
	Effect & Shift ($10^{-18}$) & Uncertainty ($10^{-18}$) \\
	\hline
	Thermal time dilation  & \textit{-0.7} & \textit{0.4} (axial) \\
                            & \textit{-2.0} & \textit{1.2} (radial) \\
                            & -2.7 (total) & 1.6 (total)\\
	\hline
	Black-body radiation & & \textit{1.3} ($\Delta\alpha_\mathrm{stat}$) \\
	& & \textit{0.2} ($T$) \\
	& -13.4 & 1.4 (total) \\
	\hline
	Quadratic Zeeman (DC) &  & \textit{1.1} ($g_e$) \\
	& & $\mathit{0.03/\sqrt{\tau}}$ ($\Delta\nu_\mathrm{Z1}$) \\
	& -35.9  & 1.1  \\
	\hline
	Time dilation (EMM) & \textit{-0.76} & \textit{0.1} (axial) \\
	& \textit{-0.05} & \textit{0.03} (radial) \\
	& -0.8 (total) & 0.1 (total) \\
	\hline
	Servo error & -2.6 & 0.5 \\
	\hline
	Electric quadrupole & & \textit{0.001} ($\omega_\mathrm{ax}$)\\
	& & \textit{0.026} ($\Theta$) \\
	& -0.14 & 0.03 (total) \\
	\hline
	Collision shift & 0 & 0.4 \\
	\hline
 	AOM chirp & -0.5 & 0.5 \\
	\hline
	First order Doppler & 0 & 0.1 \\
    \hline
    Probe laser ac Stark & 0 & 0.001 \\
	\hline
	\textbf{Total} & \textbf{-56.0} & \textbf{2.5} \\ \hline
	\end{tabularx}
\end{table}

\subsection{Time dilation shifts}
Time dilation results in a fractional frequency shift of
\begin{equation}
\left\langle\frac{\Delta\nu_\mathrm{TD}}{\nu_0}\right\rangle=-\frac{\left\langle E_\mathrm{kin}\right\rangle}{mc^2}\textnormal{,}
\end{equation}
where the kinetic energy $E_\mathrm{kin}$ contains contributions from thermal (secular) motion, intrinsic (IMM) and excess micromotion (EMM).

\subsubsection{Time dilation shift due to thermal motion}

After Doppler cooling, all $12$ motional modes $\alpha$ of the four-ion crystal are in thermal states at the respective temperatures $T_\alpha$. The associated kinetic energy consists of two contributions: Velocities of thermal motion are distributed among the ions $i$ according to the mode eigenvector components $\beta_{\alpha,i}^{\prime}$ \cite{Morigi2001}. As radial thermal excitation moves the ions into regions with nonzero rf electric field amplitude, there is an additional contribution of comparable magnitude due to intrinsic micromotion (IMM). Overall, we get \cite{Keller2019}

\begin{equation}\label{eq:TDthermal}
	\begin{split}
		\left\langle\frac{\Delta\nu_\mathrm{TD, thermal}}{\nu_0}\right\rangle_i &= -\frac{\langle v_i^2\rangle}{2c^2}=-\sum_\alpha \frac{k_BT_{\alpha}}{2c^2} \times\\\ &\times\frac{\beta_{\alpha,i}^{\prime2}}{m_i}\left(1+\frac{q_i^2}{8}\frac{\Omega_\mathrm{rf}^2}{\omega_\alpha^2}+\frac{q_i^2}{8}\right)\;\textnormal{,}
	\end{split}
\end{equation}

where $\omega_\alpha$ the eigenfrequency of mode $\alpha$, $q_i$ the Mathieu q-parameter of ion $i$ of the ponderomotive confinement by the trapping field at frequency $\Omega_\mathrm{rf}$ ($q=0$ for axial motion).

\begin{figure}
	\centering
	\includegraphics[width=.8\columnwidth]{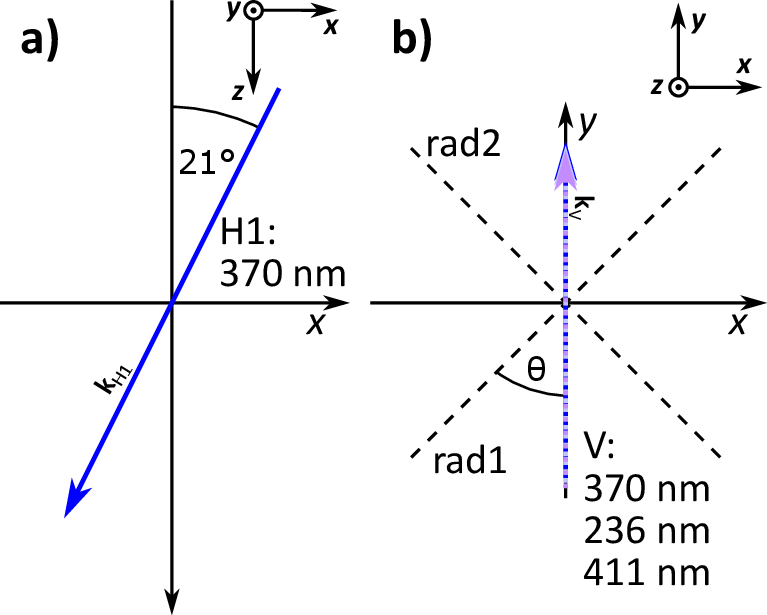}
	\caption{Cooling and clock laser geometry. Yb$^+$ Doppler cooling uses beams along two directions, termed ``H1'' and ``V''. Spectroscopic measurements with beams along ``V'' are used to determine temperature bounds.}
 \label{fig:laser_geometry}
\end{figure}

We determine upper and lower bounds for all mode temperatures by modeling the sympathetic cooling process and with spectroscopic measurements. Figure \ref{fig:laser_geometry} shows the relevant laser beam and motional mode geometry.  Since both sympathetic cooling efficiency and time-dilation impact scale with $C_\alpha=\sum_{i}\beta_{\alpha,i}^{2\prime}$, where the sum includes all cooling or clock ions, respectively, the two are inversely related. While both species participate appreciably in all axial modes, radial In$^+$ motion is predominantly restricted to a single mode ($C_{\mathrm{rad}n,4}=\unit[96]{\%}$) for each of the principal axes $\mathrm{rad}1$ and $\mathrm{rad}2$. The temperature of these modes is therefore the main concern in our choice of cooling parameters. Cooling consists of a \unit[25]{ms} pulse with a detuning of $-\Gamma/2$ and intensities of $I_\mathrm{H1}=0.3I_\mathrm{sat}$ and $I_\mathrm{V}=I_\mathrm{sat}$ in the respective beams (see Fig.~\ref{fig:laser_geometry}), followed by a \unit[25]{ms} linear ramp to $I_\mathrm{V}=0.1I_\mathrm{sat}=I_\mathrm{V,end}$ and $I_\mathrm{H1}=0.03I_\mathrm{sat}=I_\mathrm{H1,end}$. The lower bound for all 12 mode temperatures is determined by the geometry-dependent Doppler limit at $I_\mathrm{H1/V}=I_\mathrm{H1/V,end}$. For an estimate of the actual temperatures after a finite cooling pulse, we apply a linear Doppler cooling model \cite{Leibfried2003} to the entire clock cycle pulse sequence.

\begin{figure}
	\centering
	\includegraphics[width=\columnwidth]{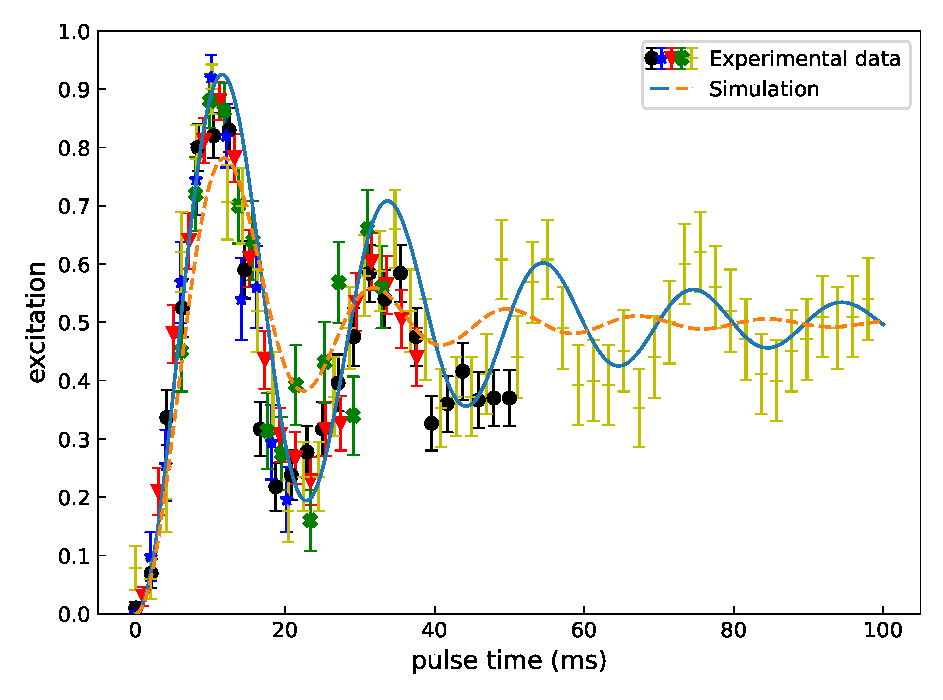}
	\caption{Thermal dephasing of Rabi oscillations observed in experiments compared to theory. Calculations assume an extremal angle $\theta_1 = \unit[27]{^\circ}$, minimizing the sensitivity to the temperature along principal axis $\mathrm{rad}2$ for a worst-case estimate. Experimental data were acquired on different days during the campaign. The solid blue line assumes the respective Doppler temperatures for mode 4 in each radial direction ($T_\mathrm{4, rad1}=\unit[0.35]{mK}$, $T_\mathrm{4, rad2}=\unit[0.65]{mK}$). The dashed orange line assumes Doppler temperature for direction $\mathrm{rad}1$ and a temperature of $T_\mathrm{4, rad2}=\unit[3.33]{mK}$ for mode 4 in direction $\mathrm{rad}2$.}
 \label{fig:Rabi_flops}
\end{figure}

Since the clock laser wave vector has nonzero projections onto both radial principal axes, thermal dephasing of Rabi oscillations contains information on the radial mode temperatures. The main uncertainty in this measurement results from our limited knowledge of the angle $\theta_1=\unit[40(13)]{^\circ}$, currently determined via spectroscopy of the \unit[411]{nm} $^2S_{1/2}\leftrightarrow{}^2D_{5/2}$ line in Yb$^+$ with limited resolution due to laser phase noise \cite{Kulosa2023}. Ground-state cooling of In$^+$ will enable a more precise determination in the future. For reasons stated above, we concentrate our analysis on the two radial modes with predominant In$^+$ motion, $\mathrm{rad}1, 4$ and $\mathrm{rad}2,4$. Due to the near-degeneracy (within \unit[2.5]{\%}) of the radial confinement, the IMM contributions according to Eq.~\ref{eq:TDthermal} are similar, and the sum of their temperatures becomes the quantity of interest. For an upper bound, we assume $\theta_1=\unit[27]{^\circ}$ -- which minimizes the sensitivity to the temperature of modes along direction $\mathrm{rad}2$ -- and the highest sum of radial temperatures compatible with our observations. Figure \ref{fig:Rabi_flops} shows data measured on different days during the campaign, together with theoretical expectations for these upper and lower temperature bounds.

Since no spectroscopic observations with sensitivity to the axial temperatures are available, we rely on our cooling calculations to estimate the respective mode temperatures. As an upper bound, we multiply these temperatures by a factor of $4.0$, which corresponds to the difference between the radial theoretical expectation and experimental upper bound. To reflect this assumed correlation, we add the uncertainties for both directions linearly. Table \ref{tab:temp_overview} summarizes the temperature bounds for all motional modes.

\begin{table*}[t]
		\begin{tabularx}{2\columnwidth}{|l|X|X|X|c|}
			\hline
			\textbf{Direction $k$} & \textbf{$T_{1,k}$} & \textbf{$T_{2,k}$} & \textbf{$T_{3,k}$} & \textbf{$T_{4,k}$} \\ \hline
			$\textbf{rad}_1$ & \multicolumn{3}{|c|}{\multirow{2}{*}{$\unit[0.82]{mK} \leq T_{1-3,\text{rad}_1}+T_{1-3,\text{rad}_4} \leq \unit[1.00]{mK}$}} & \multirow{2}{*}{$\unit[0.82]{mK} \leq T_{4,\text{rad}_1}+T_{4,\text{rad}_2} \leq \unit[3.7]{mK}$} \\ \cline{1-1}
			$\textbf{rad}_2$ &  \multicolumn{3}{|c|}{} &  \\  \hline
			$\textbf{axial}$ & \multicolumn{4}{|c|}{$\unit[0.69]{mK} \leq T_{1-4,\text{ax}} \leq \unit[2.78]{mK}$} \\ \hline
		\end{tabularx}
		\caption{Determined bounds for the temperature $T_{\text{$\alpha$,k}}$ for mode $\alpha$ in direction $k$ for the indium ion from comparison of experimental data and simulation.}
		\label{tab:temp_overview}
\end{table*}

This yields a total shift of 

\begin{equation}\label{eq:TDthermal_result}
	\left\langle\frac{\Delta\nu_\mathrm{TD, thermal}}{\nu_0}\right\rangle_\mathrm{tot}=-2.7(16)\times10^{-18}.
\end{equation}

\subsubsection{Excess micromotion}
Stray fields are compensated at least once every eight hours by discrete photon correlation measurements with three linearly independent \unit[369.5]{nm} beams and adjustments to the trap electrode voltages. We assume a linear buildup of the DC stray field during the period between these measurements and assign the time average of the resulting quadratic rise of the associated TD shift to the interval, taking the half span of DC fields as its uncertainty. This yields a radial EMM TD shift contribution of $\Delta\nu_\mathrm{EMM,rad}/\nu_0=-5(2)\times10^{-20}$. The axial rf field amplitude is caused by imperfections of the trap and has a constant value of $E_{\mathrm{rf },z} = \unit[-86.0(52)]{V/m}$. The associated time dilation shift of  $\Delta\nu_\mathrm{EMM,ax}/\nu_0=-7.6(9)\times10^{-19}$ dominates the overall contribution of $\Delta\nu_\mathrm{TD, EMM}/\nu_0=-8(1)\times10^{-19}$.

\subsection{AC Stark shifts}

\subsubsection{Black-body radiation}
The ac Stark shift due to black-body radiation (BBR) corresponding to a temperature $T$ is given by

\begin{align}\label{eq:BBR}
		h\Delta\nu_\mathrm{BBR} &=-\frac{1}{2}\left\langle E^2\right\rangle\Delta\alpha_\mathrm{stat}\left(1+\eta\right)\quad \\\ \textnormal{with}&\quad\left\langle E^2\right\rangle\approx\left(\unit[832]{V/m}\right)^2\times\left(T/\unit[300]{K}\right)^4\,\textnormal{,}
\end{align}	

where $E$ is the BBR electric field and $\Delta\alpha_\mathrm{stat} = 3.3(3)\times10^{-41}\unit{J/(V/m)^2}$  is the static differential polarizability of the clock transition \cite{Safronova2011}. Since all electric dipole transitions coupling to either of the clock states are in the UV range, the dynamical correction $\eta$ is negligible in In$^+$.

We record the environmental temperature with two Pt100 sensors on the trap wafers and four on the vacuum chamber. Fluctuations with a span of ca.~\unit[1]{K} are caused by the room temperature instability and appear on all sensors. Trap and chamber temperatures are homogeneous to within \unit[0.2]{K}, with a constant offset of ca.~\unit[1.5]{K} for the trap sensors, consistent with the expected heating by the \unit[1.2]{kV} rf trapping voltage \cite{Nordmann2020}. Averaging over all sensor readings during uptimes with equal weights and taking half the span of readings as the uncertainty, we obtain $T=\unit[299(1)]{K}$. This very coarse treatment translates to a temperature related BBR shift uncertainty of $2\times10^{-19}$, which is negligible compared to the $1.3\times10^{-18}$ contribution from the uncertainty of $\Delta\alpha_\mathrm{stat}$. In total, the shift is determined to be $\Delta\nu_\mathrm{BBR}/\nu_0=-1.34(14)\times10^{-17}$.

\subsubsection{Trap rf electric field}
Owing to the low differential polarizability of the clock transition \cite{Safronova2011}, the ac Stark shift due to $\left\langle E_\mathrm{rf}^2\right\rangle$ from EMM and IMM is more than an order of magnitude below the corresponding TD shift.

\subsubsection{Probe-field-induced ac Stark shift}
We have assessed two types of shift from off-resonant coupling of the probe light: to other Zeeman components \cite{Yudin2023} and to other electronic transitions. In frequency scans at increased power, we could not observe any excitation of unwanted Zeeman components and thus determine an upper bound on polarization impurity and magnetic field misalignment which correspond to a shift of $4\times10^{-22}$ \cite{vonBoehn2022}. Summing over all relevant electric dipole transitions to other electronic states yields an upper bound for their effect on the clock transition frequency of $7\times10^{-22}$. All other laser beams are blocked by mechanical shutters during interrogation.

\subsection{Second-order Zeeman shift}
The second-order Zeeman sensitivity of the $\ket{^3P_0,m_F=\pm9/2}$ states can be written as
\begin{equation}\label{eq:2nd_order_zeeman}
	\Delta\nu_\mathrm{Z2}=-\frac{2\mu^2_B}{3h^2\Delta_\mathrm{FS}}\langle B^2\rangle=\unit[-4.05]{Hz/mT^2}\times\langle B^2\rangle\textnormal{,}
\end{equation}
where $\mu_B$ is the Bohr magneton and $\Delta_\mathrm{FS}=\unit[32.22]{THz}$ \cite{A.Kramida2021} is the fine-structure splitting between $^3P_0$ and $^3P_1$. This expression takes the dominant coupling between $^3P_0$ and $^3P_1$ into account. A small correction is expected from the contribution of the $^1P_1$ level. We determine the magnetic field $B$ during clock operation from the first-order Zeeman splitting as
\begin{equation}
	B=\frac{\nu_+ - \nu_-}{2\left(\vert m_e\vert g_e - \vert m_g\vert g_g\right)\mu_B/h}=\frac{\nu_+ - \nu_-}{2m\left(g_e - g_g\right)\mu_B/h}\;\textnormal{,}
\end{equation}
where $m=\vert m_g\vert=\vert m_e\vert=9/2$, and $\nu_\pm$ denotes the frequency of the transition  $\ket{^1S_0, m_F=\pm9/2} \leftrightarrow \ket{^3P_0, m_F=\pm9/2}$. The relative uncertainty of this measurement is limited to $\unit[1.5]{\%}$ by the uncertainty of $g(^3P_0)-g(^1S_0)=-3.21(5)$ \cite{Becker2001}, which translates to a $\unit[3]{\%}$ uncertainty of the second-order shift. Compared to this, the contribution of the frequency splitting uncertainty of $\sigma(\nu_+-\nu_-)/(\nu_+-\nu_-)=4\times10^{-4}/\sqrt{\tau/\unit[1]{s}}$ is negligible. We apply these corrections with a time resolution of about \unit[14]{s} to the frequency data in post-processing. The mean shift during clock uptime is $\Delta\nu_\mathrm{Z2}/\nu_0=-3.59(11)\times10^{-17}$.

From a model of the rf currents in our trap, we expect an rms ac magnetic field due to the trap drive of $B^2_{\text{rms}}=1.3\times10^{-12}T^2$ \cite{Dolezal}, which would correspond to a shift of $-9\times10^{-21}$.

\subsection{Servo Error}

Frequency corrections are determined from measurements at four laser detunings: For each of the two Zeeman components, we detune the clock laser by $\delta\nu=\pm\unit[3.3]{Hz}$. Integrating servos keep track of the center frequency and splitting of the two transitions. The algorithm includes second-order corrections in the discrete updates to avoid an offset due to the linear drift of the Si cavity \cite{Peik2005}. However, due to a software bug at the time, the applied corrections were about a factor of $4$ too small.

To account for this residual offset, we reconstruct the error signal from logged excitation raw data and convert it to frequency errors using a slope of $\unit[0.153(17)]{Hz^{-1}}$, determined from frequency scans. The mean of this signal corresponds to the servo error, except for an additional offset in the presence of asymmetrically distributed dead time within each update interval. We infer a drift rate of $\unit[-380]{\mu Hz/s}$ (for the frequency-quadrupled radiation at \unit[1.267]{THz}) from the servo corrections and assume up to $\pm\unit[1]{s}$ of dead-time asymmetry per update, which contributes an uncertainty of $\unit[380]{\mu Hz}=3\times10^{-19}\times\nu_0$. Overall, this treatment yields a correction of $\Delta\nu_\mathrm{servo}/\nu_0=-2.6(5)\times10^{-18}$, with the uncertainty determined by the statistical uncertainty of the signal distribution and the parameter uncertainties and dead-time asymmetry contributions stated above. A continuous correction of the first-order drift will eliminate these contributions in future campaigns.

\subsection{Collision shift}
\begin{figure}
	\centering
	\includegraphics[width=\columnwidth]{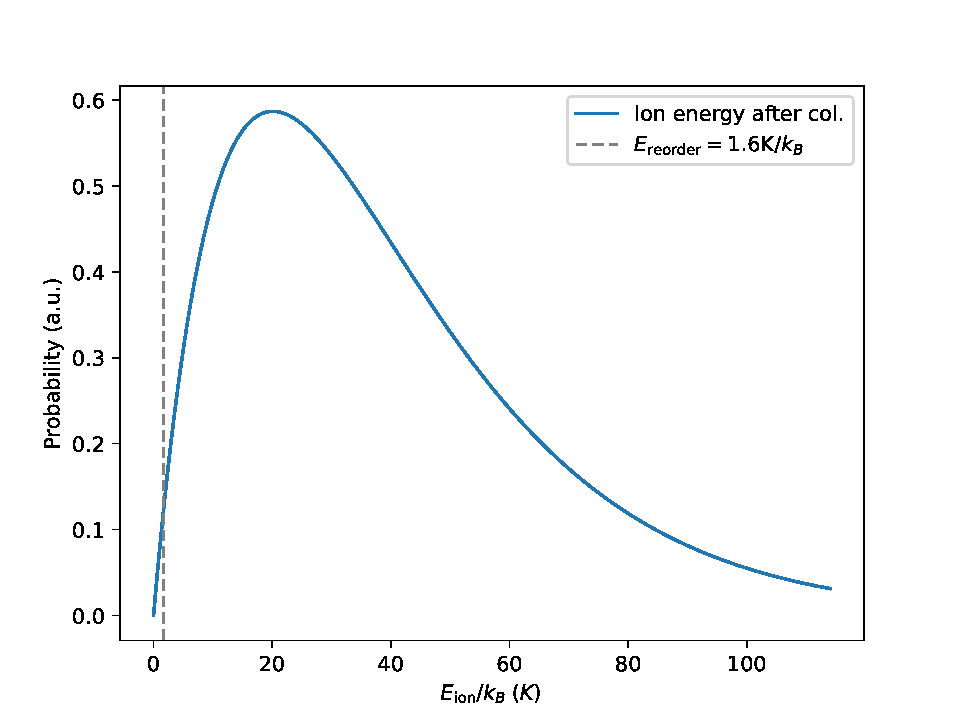}
	\caption{Energy distribution of indium ions after collision with H$_2$ background gas molecules at a thermal distribution with $T=\unit[299]{K}$. The reorder energy $E_\mathrm{reorder}=k_B\unit[1.6]{K}$ (dashed gray line) is calculated analogous to Ref. \cite{Hankin2019} with the secular frequencies of ($\omega_{\text{rad1}}, \omega_{\text{rad2}}, \omega_{\text{ax}})/2\pi\approx(\unit[1.23]{MHz}, \unit[1.19]{MHz},\unit[336]{kHz})$ ($^{115}$In$^+$ center-of-mass motion).}
 \label{fig:collisions}
\end{figure}

We determine the rate of collisions above the ion swapping energy barrier to be $\Gamma_{\text{meas}}^{\text{ion}}=\unit[0.0029(3)]{s^{-1}}$ per ion by observing swap events in mixed-species chains of different lengths  \cite{Nordmann2023a}. The derivation of this rate assumes that all crystal permutations are equally likely to occur after a collision and accounts for the likeliness of encountering indistinguishable permutations (in which each position is occupied by an ion of the same species before and after the event). Assuming a background gas of $H_2$ molecules at \unit[299]{K}, the fraction of collisions with energy transfer below this barrier is negligible (see Fig.~\ref{fig:collisions}). Our system detects all collisions that lead to detectable permutation changes, which corresponds to \unit[75]{\%} of events for the 1In$^+$-3Yb$^+$ crystal. The rate of undetected collisions with any ion in the crystal is thus $\Gamma_{\text{undet}}^{\text{CC}}=\unit[25]{\%}\times4\times\Gamma_{\text{meas}}^{\text{ion}}=\unit[0.0029(3)]{s^{-1}}$. Using this rate in the treatment described in Ref.~\cite{Rosenband2008} yields a collision shift of $\Delta\nu_\mathrm{coll}/\nu_0 = 0(4)\times10^{-19}$ due to phase shifts of the electronic states. Any TD shifts of comparable magnitude would be subject to the suppression mechanisms described in Ref.~\cite{Hankin2019}.

\subsection{Electric quadrupole shift}
Electric quadrupole (E2) shifts are caused by electric field gradients coupling to the quadrupole moment of the $^3P_0$ state of $\Theta=-1.6(3)\times10^{-5}ea_0^2$ \cite{Beloy2017}, where $e$ is the elementary charge and $a_0$ the Bohr radius. Gradients are caused by the static components of the confining potential, the charges of the other ions in the chain, and potentially by external stray electrostatic potentials. For the stretched Zeeman states ($m_F=\pm F$) and a magnetic field alignment along the trap axis, the trap and CC contributions for ion $i$ are \cite{Itano2000, Keller2019}
\begin{align}\label{eq:E2shift}
	\left\langle\frac{\Delta\nu_{E2}}{\nu_0}\right\rangle_i &=\frac{1}{h\nu_0}\Theta\frac{m\omega_\mathrm{ax}^2}{e} \times \\\ &\times\left(\underbrace{\frac{1}{2}}_{\mathrm{trap}}+\underbrace{\sum_{j=1;j\neq i}^N\frac{1}{\vert u_i-u_j\vert^3}}_{\mathrm{other\: ions}}\right)\;\textnormal{,}
\end{align}
where $u_i$ are scaled equilibrium positions of the ions in a harmonically confined chain \cite{James1998}, and  $\omega_\mathrm{ax} =2\pi\times \nu_\mathrm{ax, In}=2\pi\times\unit[336.3(12)]{kHz}$. For the In$^+$ ion in one of the inner positions of a four-ion chain, this results in a fractional shift of $\Delta\nu_\mathrm{E2}/\nu_0=-1.4(3)\times10^{-19}$, with the dominating uncertainty contributed by $\Theta$. The trap frequency uncertainty contributes $1\times10^{-21}$ and the shift is first-order insensitive to magnetic field pointing deviations in our configuration. We derive an upper bound for the contribution of stray field gradients by assuming that the stray fields observed in EMM compensation originate from a point charge on one of the respective electrode surfaces. These are the nearest surfaces to the equilibrium positions of the ions. The obtained fractional frequency shifts are below $10^{-21}$ for the entire duration of the measurements and are therefore neglected.

\subsection{First-order Doppler shifts}

Net optical path length changes between the clock laser light at the frequency comb and ions during the measurements, e.g., due to cycle-synchronous mechanisms, will result in first-order Doppler shifts. In this section, we give estimates of multiple possible contributions.

\subsubsection{Optical path length fluctuations}
We use active phase stabilization servos \cite{Ma1994} for the fibers from the laser table to the ULE cavity, frequency comb, and vacuum chamber tables. At the frequency comb, all optical paths are end-to-end stabilized \cite{Benkler2019}. The in-loop beat frequencies of all path length stabilizations are monitored with frequency counters. Data are rejected whenever deviation from the target frequency indicates a cycle slip.

Thermal expansion of the optical table in an unstabilized section close to the ion trap contributes a first-order Doppler shift of 

\begin{equation}\label{eq:first_order_Doppler_table}
	\frac{\Delta\nu_\mathrm{1D}}{\nu_0} = \frac{v}{c} = \frac{1}{c} \frac{dl}{dT} \frac{dT}{dt} = \frac{l\alpha}{c} \frac{dT}{dt}\mathrm{,}
\end{equation}

with $\alpha=\unit[20(5)]{\mu m /(m\cdot K)}$ the coefficient of thermal expansion of steel, $l$ the unstabilized path length and $c$ the speed of light. The optical table temperature is determined from a nearby room temperature sensor and low-pass filtered via a rolling average over \unit[1800]{s} to suppress readout noise. Averaging this shift over the clock uptime, we obtain an overall contribution with a magnitude of $1\times10^{-19}$ for a total unstabilized path length of about \unit[5]{m}.

\subsubsection{UV-induced stray electric fields}
Photo-electrons released by the UV lasers can accumulate on dielectric surfaces and shift the ion equilibrium positions synchronously with the interrogation cycle. We investigate this effect by monitoring the micromotion amplitude as a proxy for ion position changes while applying light from either the clock or detection laser. In the radial plane, the rf electric field amplitude and ion displacement are related by 
\begin{equation}\label{eq:displ_ion_Erf_2D}
	\Delta x/y=\frac{2e}{qm\Omega_{\textrm{rf}^2}}E_{\textrm{rf,y/x}}\textrm{,}
\end{equation}
where $q$ is the Mathieu q-parameter for the ponderomotive confinement. Note that an ion displaced in the $x$ direction is subjected to an rf field along the $y$ direction and vice-versa in our reference system.

During continuous illumination with clock laser light at about $100$ times the power used during clock operation, we observe changes of the equilibrium position of a single Yb$^+$ ion which follow an exponential trajectory with a time constant of about \unit[470]{s}. While the initial velocity of this movement would correspond to a Doppler shift of about $3\times10^{-18}$, this motion is orthogonal to the clock laser k-vector and therefore does not affect the interrogation. No movement of the ion has been observed in response to illumination with the \unit[230.6]{nm} detection laser light.

\subsubsection{Acousto-optical modulator phase chirp}

The clock laser light is switched using an AOM. Thermal effects in the crystal can cause cycle-synchronous optical path length fluctuations and a corresponding Doppler shift. We have measured the clock laser phase interferometrically with the device described in Ref. \cite{Kazda2016} for different RF drive powers $P_\mathrm{rf}$ of the AOM, using the same duty cycle as during clock operation, with $t_{\mathrm{on}}=\unit[150]{ms}$ and $t_{\mathrm{off}}=\unit[180]{ms}$. We observe a linear relation between the frequency shift and drive power: $\Delta\nu_\mathrm{AOM}/P_\mathrm{rf} = \unit[-0.11(1)]{mHz/mW}$. Since $P_\mathrm{rf}$ is adjusted to stabilize the optical power close to the ion trap, we assume a range $\unit[0]{mW}<P_\mathrm{rf}<\unit[11]{mW}$, resulting in shifts of $-1\times10^{-18}<\Delta\nu_\mathrm{AOM}/\nu_0<0$. The applied correction and its uncertainty therefore cover this entire range, $\Delta\nu_\mathrm{AOM}/\nu_0=-5(5)\times10^{-19}$ .

\subsection{Relativistic redshift}

The relative heights of all involved clocks, as well as the local gravitational acceleration, were determined in local leveling campaigns on the PTB campus in order to derive the corresponding relativistic redshift corrections \cite{Denker2018} (sometimes referred to as ``gravitational redshift''). Tab.~\ref{tab:rrs_overview} lists these corrections for the indium clock with respect to the other clocks.

\begin{table}
	\caption{Fractional relativistic redshift (RRS) corrections of the indium Coulomb crystal clock in the reference frame of the respective clock.}
	\begin{tabularx}{\columnwidth}[t]{|X|X|}
		\hline
		\textbf{Reference clock} & \textbf{RRS correction ($10^{-16}$)} \\
		\hline
		$^{171}$Yb$^+$ & 3.259(5) \\
		$^{87}$Sr & 4.135(5) \\
		$^{133}$Cs & 3.832(7) \\
		\hline
	\end{tabularx}
	\label{tab:rrs_overview}
\end{table}

\section{$\chi^2$-tests of the ratio measurements}
\begin{figure}
  \centerline{\includegraphics[width=.5\textwidth]{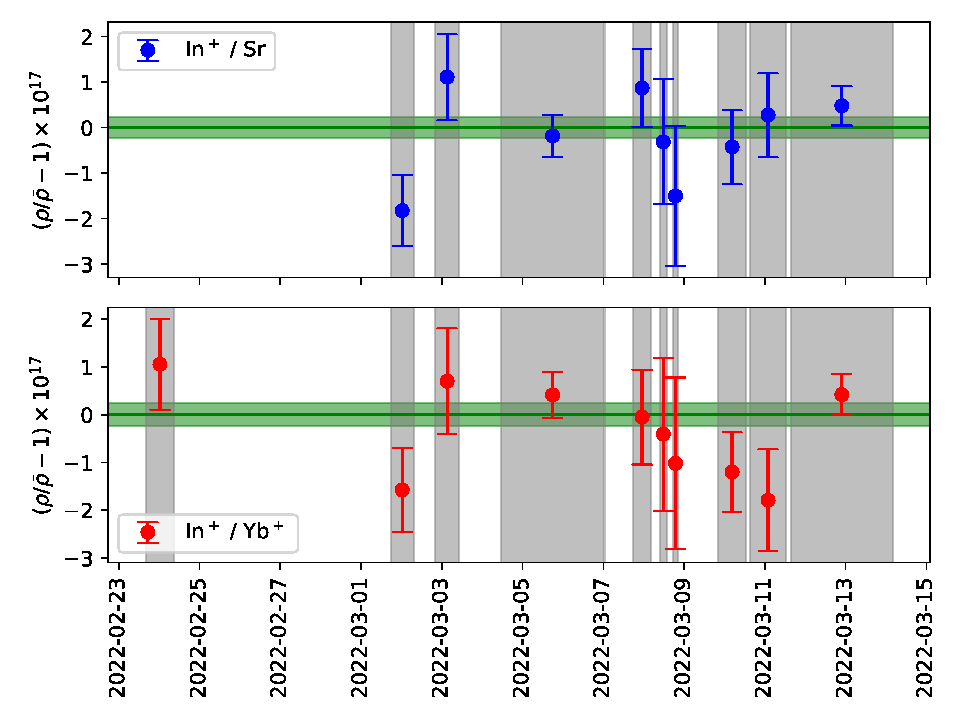}}
  \caption{\label{fig:ratio_time_series} Time series of the measured ratios. The intervals correspond to runs of the In${}^+$ clock. All depicted uncertainties are purely statistical.}
\end{figure}

We test the consistency of our observations with the inferred statistical uncertainties via $\chi^2$-tests of time series data for both optical frequency ratios. Figure \ref{fig:ratio_time_series} shows a separation into runs of the In${}^+$ clock. Each interval is assigned a mean value $\rho_i$ with a statistical uncertainty of $\sigma_i = \sigma_0/\sqrt{t_i / \unit[1]{s}}$, where $t_i$ is the duration of valid data (uptime overlap of both clocks) and $\sigma_0$ is the extrapolation to \unit[1]{s} averaging time of the fitted level of white frequency noise observed in the full dataset of the respective ratio (see Fig. 2 of the main text). We calculate a weighted mean as $\bar{\rho}=\sum_iw_i\rho_i/\left(\sum_iw_i\right)$, where $w_i=\sigma_i^{-2}$, and the corresponding $\chi^2_\mathrm{red}=\frac{1}{N-1}\sum_{i=1}^Nw_i(\rho_i-\bar{\rho})^2$, and obtain $\chi^2_\mathrm{red}=1.33$ (In${}^+$/Sr) and $\chi^2_\mathrm{red}=1.30$ (In${}^+$/Yb${}^+$).

\section{Repeated comparison and additional correction of the Sr clock frequency}

\begin{figure}
  \centerline{\includegraphics[width=.4\textwidth]{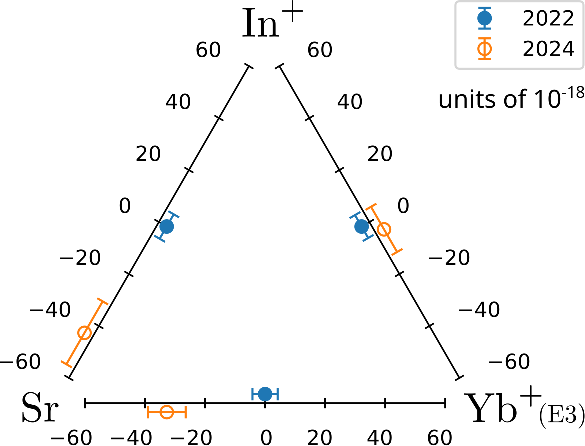}}
  \caption{\label{fig:ratios_2022_2024}
  Visualization of the optical frequency ratio measurements in 2022 and 2024, displayed as fractional changes with respect to the 2022 values in units of $10^{-18}$. All ratios are defined with the higher frequency in the numerator. The deviations are consistent with an increase of the Sr clock frequency. The values correspond to those listed in Tab.~\ref{tab:ratios_2022_2024}.
  }
\end{figure}

A frequency ratio of $\nu^{\text{In}^+}_0/\nu^{\text{Sr}}_{0}=2.952\,748\,749\,874\,860\,909(14)$ has been measured directly, where the uncertainty includes the statistical contribution and the contributions from known systematic corrections.
In particular, the systematic uncertainty of the strontium clock has been estimated to be $3.2 \times 10^{-18}$ using the procedures described in Ref.~\cite{Schwarz2022}.
However, the frequency ratio $\nu^{\text{Yb}^+}_0/\nu^{\text{Sr}}_{0}$ observed with respect to the ytterbium ion clock is larger for this apparatus, Sr3 \cite{Schwarz2022}, than for its predecessor, Sr1 \cite{Doerscher2021}. In a repeated comparison between Sr3 and the In$^+$ and Yb$^+$ (E3) clocks in 2024, we found deviations which we attribute to the strontium clock (see Fig.~\ref{fig:ratios_2022_2024} and Tab.~\ref{tab:ratios_2022_2024}).
These observations have raised concern about an unidentified systematic frequency shift at the $10^{-17}$ level.
Investigations into this matter are on-going at the time of writing.

\begin{table*}
\begin{tabular}{llll}
\textbf{Clocks} & \textbf{Freq.\ ratio 2022} & \textbf{Freq.\ ratio 2024} & \textbf{Frac. difference} $\times10^{18}$\tabularnewline\hline
    $\mathrm{In}^+$ / $\mathrm{Yb}^+$ & 1.973773591557215789(9) & 1.973773591557215780(17) & $-5(10)$\\
    $\mathrm{In}^+$ / $\mathrm{Sr}$ & 2.952748749874860909(14) & 2.952748749874860778(36) & $-44(13)$\\
    $\mathrm{Yb}^+$ / $\mathrm{Sr}$ & 1.495991618544900664(6) & 1.495991618544900615(9) & $-33(8)$\\
\end{tabular}
\caption{\label{tab:ratios_2022_2024} Frequency ratio values of the clocks used in this work, measured in March 2022 and March 2024.}
\end{table*}

In the remainder of this section, we derive a provisional frequency correction estimate to allow comparison with the ratio $\nu^{\text{In}^+}_0/\nu^{\text{Sr}}_{0}$ published by NICT \cite{Ohtsubo2020}. However, we only consider the measurements with respect to Yb$^+$ and Cs as metrologically relevant results for a future evaluation of the In$^+$ recommended frequency \cite{Margolis2024}.

\begin{figure}
  \centerline{\includegraphics[width=.5\textwidth]{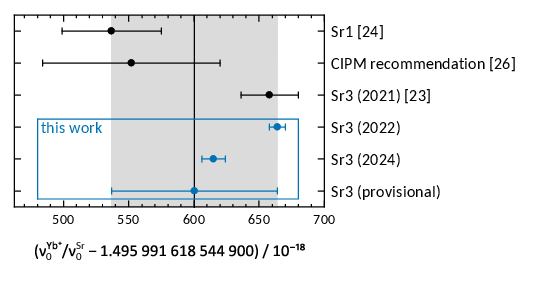}}
  \caption{\label{fig:yb_sr}
  Measurements of the optical frequency ratio $\nu^{\text{Yb}^+}_0/\nu^{\text{Sr}}_{0}$.
  In addition to the values reported for the Sr1 clock \cite{Doerscher2021}, recommended by the CIPM \cite{Margolis2024}, and reported earlier for the Sr3 clock \cite{Schwarz2022}, two results measured in this work are shown.
  The latter were used to attribute excess scatter to the Sr3 clock, see Fig.~\ref{fig:ratios_2022_2024}.
  The estimate of the frequency ratio that is used to derive a provisional correction for the Sr3 clock in the text is shown in the bottom row; its value and uncertainty are also indicated by the vertical line and shaded area, respectively.
  }
\end{figure}

In order to find bounds for the frequency shift of Sr3 from the unperturbed $\nu^{\text{Sr}}_{0}$, we compare the frequency ratios measured in this work \footnote{Here, the systematic uncertainty once again includes only the contributions from known effects.} to other determinations with uncertainties below $10^{-16}$
\footnote{
The frequency ratio reported in Ref.~\cite{Doerscher2021} results from about a hundred individual measurements using the Sr1 clock, including a few where the frequency ratio was almost as large as measured using Sr3 in this work.
For the sake of simplicity and to facilitate the assessment of correlations, we only consider the final frequency ratio reported in Ref.~\cite{Doerscher2021}.
}, as shown in Fig.~\ref{fig:yb_sr}.
From the available data, we estimate that the true value is likely included in the interval of the observations, which is bounded by $\nu^{\text{Yb}^+}_0/\nu^{\text{Sr}}_{0} = 1.495\,991\,618\,544\,900\,537(38)$ \cite{Doerscher2021} and the value observed in this work in 2022.
We thus correct the strontium clock's frequency in this work to the center of the interval with an uncertainty equal to half the interval's width (see Fig.~\ref{fig:yb_sr}), which corresponds to a fractional correction of $+42(42) \times 10^{-18}$.
This results in the corrected frequency ratio $\nu^{\text{In}^+}_0/\nu^{\text{Sr}}_{0}=2.952\,748\,749\,874\,860\,78(13)$ given in the main text.
Note that this treatment introduces a substantial correlation between the reported final $\nu^{\text{In}^+}_0/\nu^{\text{Sr}}_{0}$ ratio and the $\nu^{\text{Yb}^+}_0/\nu^{\text{Sr}}_{0}$ values used to derive the correction.

\section{Correlation coefficients}
Various quantities reported in this work show a degree of correlation with respect to each other and to existing literature due to the participation of the same clock in two comparisons, the use of the same systematic correction evaluation, etc. In Tab.~\ref{tab:correlations}, we quantify the most relevant of these coefficients, following the definitions as stated in the corresponding guidelines, which have been assembled within the EMPIR 18SIB05 ROCIT project \cite{RocitCorrelations2020}. We do not state correlations for measurements involving Sr3, which we do not consider relevant for the update of recommended frequencies due to the provisional frequency correction derived above.

\onecolumngrid

\begin{table}[h]
    \begin{tabularx}{1\columnwidth}{p{0.3\columnwidth}p{0.3\columnwidth}p{0.3\columnwidth}p{0.1\columnwidth}}
    \textbf{$q_i$} & \textbf{$q_j$} & \textbf{Correlated quantity} & \textbf{$r(q_i,q_j)$}\\
    \hline
    $\nu^{\text{In}^+}_0/\nu^{\text{Yb}^+}_0$ (this work) & $\nu^{\text{In}^+}_0$ (this work, via Cs) & In$^+$ sys. unc. & $0.01$\\
    $\nu^{\text{In}^+}_0$ (this work, via Yb$^+$) & $\nu^{\text{In}^+}_0$ (this work, via Cs) & In$^+$, CSF2 sys. unc. & $0.55$\\
    $\nu^{\text{In}^+}_0$ (this work, via Cs) & $\nu^{\text{Yb}^+}_0$ \cite{Lange2021} & CSF2 sys. unc. & $0.55$\\
    $\nu^{\text{In}^+}_0$ (this work, via Yb$^+$) & $\nu^{\text{Yb}^+}_0$ \cite{Lange2021} & $\nu^{\text{Yb}^+}_0$ & $1.00$\\
    \end{tabularx}
    \caption{\label{tab:correlations} Correlation coefficients between quantities reported in this work and previous literature.}
\end{table}

\end{document}